\documentclass[aps,preprint,prb,showpacs]{revtex4-1}%
\usepackage[colorlinks=true,linkcolor=blue]{hyperref}
\usepackage{amssymb}
\usepackage{amsfonts}
\usepackage{amsmath}
\usepackage{graphicx}%
\providecommand{\U}[1]{\protect\rule{.1in}{.1in}}
\begin{document}
\preprint{REV\TeX4-1 }
\title[Geometric Momentum on Two-dimensional Sphere]{Geometric Momentum: the Proper Momentum for a Free Particle on a
Two-dimensional Sphere}
\author{Q. H. Liu}
\email{quanhuiliu@gmail.com}
\affiliation{School for Theoretical Physics, and Department of Applied Physics, Hunan
University, Changsha, 410082, China}
\author{L. H. Tang}
\affiliation{School for Theoretical Physics, and Department of Applied Physics, Hunan
University, Changsha, 410082, China}
\author{D. M. Xun}
\affiliation{School for Theoretical Physics, and Department of Applied Physics, Hunan
University, Changsha, 410082, China}

\begin{abstract}
In Dirac's canonical quantization theory on systems with second-class
constraints, the commutators between the {position, momentum} and Hamiltonian
form a set of algebraic relations that are fundamental in construction of both
the quantum momentum and the Hamiltonian. For a free particle on a
two-dimensional sphere or a spherical top, results show that the well-known
canonical momentum $p_{\theta}$ breaks one of the relations, while three
components of the momentum expressed in the three-dimensional Cartesian system
of axes as $p_{i}$ $(i=1,2,3)$ are satisfactory all around. This momentum is
not only geometrically invariant but also self-adjoint, and we call it
geometric momentum. The nontrivial commutators between $p_{i}$ generate three
components of the orbital angular momentum; thus the geometric momentum is
fundamental to the angular one. We note that there are five different forms of
the geometric momentum proposed in the current literature, but only one of
them turns out to be meaningful.

\end{abstract}
\date{\today}

\maketitle

\section{Introduction}

Although the fundamental principles of quantum constrained dynamics were
outlined by Dirac in the 1950s to 1960s \cite{dirac1}, no consensus regarding
the formulation has so far been reached when applied to particular systems of
second-class constraints \cite{dispute,book1,book2,book3,book4}. A subject of
intense debate for decades has been: what the proper definition of momentum in
quantum mechanics is for a free particle on a two-dimensional sphere. To
obtain an unambiguous conclusion, we carefully analyze the geometric momentum
(GM), which will be explicitly defined shortly. From the viewpoint of the GM,
there are more than five different understandings, and each has its own
exclusive conclusions from which we can see how physics proceeds with
excitement, conflict, and controversy \cite{book2}. In this paper, we do not
extensively deal with sphere $S^{N}$ in any ($N+1)$-dimensional flat space
unless $N=2$, and this line of reasoning is applicable for sphere $S^{N}$ when
$N\geq2$.

\textit{According to Dirac, the canonical quantization procedure is an
effective recipe of quantizing a classical theory while attempting to preserve
the formal structure of the canonical Poisson brackets, or the Dirac brackets
for a constrained system, to the extent possible} \cite{dirac1,dirac2}.
However, from the very beginning, Dirac was well aware of the difficulty of
the procedure, and he commented on it in a footnote of his \textit{Principles}
\cite{dirac2}, stressing that it is fundamentally correct only in Cartesian
coordinates. Some think that the phrase \textit{Cartesian coordinates} used by
Dirac actually implies the Cartesian space, i.e.,\ the flat space
\cite{maraner,1997-1,1998,2001,2006-1,1968,1983,1985,1992,1996,2000,1990,2003,2006-2,2007,2009-1,2004,2006-3,2006-4,2006-5,2005,2008,2010-1}%
. If so, what is the quantization procedure is developed in flat space, where
the two-dimensional curved surface is embedded? Basically, investigations have
been performed from two perspectives, and ours in the present work belongs to
the second. The first is solely from the intrinsic point of view, dealing with
quantities such as the distance element square $ds^{2}=$\ $g_{\mu\upsilon
}dx^{\mu}dx^{\upsilon}$ within the surface as measured along curves on the
surface parametrized by ($x^{\mu},x^{\upsilon}$), the Laplace-Beltrami
operator $\nabla^{2}=1/\sqrt{g}\partial_{\mu}g^{\mu\upsilon}\sqrt{g}%
\partial_{\upsilon}$, and the Gaussian curvature $K$. This approach allows
mapping from the surface to a flat space, e.g., gnomonic projection
\cite{higgs}, multivalued mapping \cite{book5,multi}, etc \cite{maupertuis}.
The second is from both intrinsic and extrinsic points of view, where the
presence of the mean curvature $H$, defined via the surface's embedding in a
flat space, is a salient feature. In differential geometry, a two-dimensional
surface cannot be completely specified unless two curvatures are known. How
the constrained motion on the surface is advanced from the second prospective
in recent decades will be discussed next.

\subsection{Quantization in flat space and geometric potential and geometric
momentum}

When a particle moves on a two-dimensional curved surface, we examine it in
three-dimensional flat space rather than on the surface itself. Moreover,
because no geometric surface of zero thickness exists in the realistic world,
it is better to conceive of a surface as a limiting case of a curved shell of
equal thickness $z_{0}$, where the limit $z_{0}\rightarrow0$ is then taken.
Roughly speaking, there is a noninterchangeability of computational order:
first taking limit $z_{0}\rightarrow0$ and then defining the derivatives on
the surface and first taking derivatives in bulk and then letting
$z_{0}\rightarrow0$. This observation sheds new light on the quantization of
the kinetic energy $T=p^{2}/(2\mu)$\ and the momentum operator $\mathbf{p}$,
which are reviewed with comments in the following.

Progress on the quantization of the kinetic energy was first initiated by
Jensen, Koppe in 1971 \cite{jk}, and was finally completed by da Costa in 1981
\cite{dacosta}. Jensen, Koppe, and da Costa developed the so-called confining
potential approach in which the two-dimensional curved surface is realized by
limiting the three-dimensional system originally defined in flat space, and
they found that the correct result of the quantum kinetic energy is
\begin{equation}
T=-\frac{\hbar^{2}}{2\mu}\nabla^{2}-\frac{\hbar^{2}}{2\mu}(H^{2}-K).
\label{GP}%
\end{equation}
This is a striking finding, and it is not only a gem in quantum mechanics but
also a treasure in surface mathematics. Eq. (\ref{GP}) implies that the
original Laplace-Beltrami operator $\nabla^{2}$ on the surface may not be
enough unless a term $(H^{2}-K)$ is included,
\begin{equation}
\nabla^{2}\rightarrow\nabla^{2}+(H^{2}-K). \label{GP1}%
\end{equation}
When an electromagnetic wave transmits through thin-layer media as the
thickness is negligible, the curvature-dependent part $(H^{2}-K)$ of the
Laplace-Beltrami operator can produce observable effects \cite{2010-2}. In
quantum mechanics, the constraint-induced curvature-dependent potential
$-\hbar^{2}/(2\mu)(H^{2}-K)$ was simply called the geometric potential
\cite{2010-2}, and it is attractive for $(H^{2}-K)\geq0$. Because $H^{2}-K=0$
for the $S^{2}$ sphere, the Hamiltonian for a point particle freely moving on
the surface still assumes the usual form as $H=T=-\hbar^{2}\nabla^{2}/(2\mu)$,
where $H$ should not be confused with the mean curvature. The confining
potential procedure of quantization has been widely employed and investigated
ever since
\cite{pot1,pot2,pot3,pot4,pot5,pot6,pot7,pot8,pot9,pot10,pot11,pot12,pot13}.
From the experimental point of view, as many noted \cite{pot11}, the
realization of an optical analog of the curvature-induced geometric potential
can be taken as empirical evidence for the validity of the confining potential
procedure \cite{2010-2}.

As to the quantization of the momentum operator $\mathbf{p=(}p_{x},p_{y}%
,p_{z}\mathbf{)}$ expressed in the three-dimensional Cartesian coordinate
system $(x,y,z)$, a remarkable finding was made in 2007 \cite{2007}, and the
general form is given by
\begin{equation}
\mathbf{p}=-i\hbar(\mathbf{r}^{\mu}\partial_{\mu}+H\mathbf{n}), \label{GM}%
\end{equation}
where $\mathbf{r=[}x(x^{\mu},x^{\upsilon}),y(x^{\mu},x^{\upsilon}),z(x^{\mu
},x^{\upsilon})\mathbf{]}$ is the position vector on the surface,
$\mathbf{r}^{\mu}=g^{\mu\nu}\mathbf{r}_{\nu}=g^{\mu\nu}\partial\mathbf{r/}%
x^{\nu}$, and at this point $\mathbf{r}$, $\mathbf{n=(}n_{x},n_{y}%
,n_{z}\mathbf{)}$ denotes the normal and $H\mathbf{n}$ symbolizes the mean
curvature vector field, a geometric invariant \cite{2007}. In contrast to the
geometric potential we can call it GM because it in fact depends on the mean
curvature. If simply denoting the gradient operator $\mathbf{r}^{\mu}%
\partial_{\mu}$\ on the surface as $\nabla$, Eq. (\ref{GM}) implies the
following correspondence:
\begin{equation}
\nabla\rightarrow\nabla+H\mathbf{n.} \label{GM1}%
\end{equation}

Quantities (\ref{GP}) and (\ref{GM}) assume their usual from, provided the
surface is flat, $H=K=0$. This is good, but an immediate question arises:\ Can
these new quantities be formulated into Dirac's theory for constrained
motions? In the general case, this question is open, and we will give an
answer in the near future. This paper focuses on an elaboration of the motion
on a spherical surface where the answer is affirmative, as we see in sections
III and VI.

\subsection{Comments on the routine paradigm of quantization on the curved
surface}

The routine paradigm of quantization toward the motion on the surface is
performed within intrinsic geometry irrespective of the existence of the
higher-dimensional flat space in which the surface could be embedded. When
applied to the motion of a point particle on two-dimensional surface, it is
usually postulated \cite{podolsky,Schiff,dewitt} that the Cartesian coordinate
Laplacian should be replaced by the Laplace-Beltrami operator acting on this
manifold and the kinetic energy is $T=-\hbar^{2}\nabla^{2}/(2\mu)$ without the
geometric potential. This is highly controversial and has been debated for
decades. If the surface is a two-dimensional sphere $S^{2}$ of radius $r$,
$T=-\hbar^{2}\nabla^{2}/(2\mu)=$ $L^{2}/(2\mu r^{2})$ where $L$ is the angular
momentum of the particle. DeWitt studied the path-integral quantization of
this problem and found an extra energy term proportional to the Gaussian
curvature $1/r^{2}$ in the Hamiltonian as $\alpha\hbar^{2}/r^{2}$, with
$\alpha=1/24$ \cite{dewitt}. Others gave $\alpha=1/12$ \cite{Cheng} and
$\alpha=1/8$ \cite{Dekker}. These zero-point energies conflict with the cosmic
observation \cite{book2,kleinert} and must be discarded. Therefore, to deal
with the quantization of the kinetic energy, on one hand, there is therefore
no transparent and self-consistent way within this routine manner; on the
other, as we understand today, the absence of the geometric potential is
contradictory to the experiments. However, as to describing a free motion with
curvilinear coordinates in the flat space, Podolsky's recipe gives for the
kinetic energy $T=-\hbar^{2}\nabla^{2}/(2\mu)$ \cite{podolsky}. This is,
nevertheless, perfect. What is more, the quantum mechanics in flat space can
be rewritten into curved space with a Maupertuis metric \cite{maupertuis},
where the Laplace-Beltrami operator acquires an extra curvature term
corresponding to a conformally invariant Laplace operator. In momentum space
the quantum mechanics of the hydrogen-atom is driven by a Laplace operator
with yet another extra curvature term \cite{maupertuis}. Conversely, with the
help of a multivalued mapping it is possible to map the known flat-space
physics into the curved-space physics with only the Laplace-Beltrami operator
\cite{book5,multi}.

\subsection{Previous utilization of Dirac's canonical quantization theory}

Let us turn to the Dirac's canonical quantization theory of a system with
second-class constraints. Previous works find that various momentum including
the usual canonical one and GM as well are all definable \cite{1983,1990}. For
a two-dimensional sphere
\cite{1968,1983,1985,1992,1996,2000,1990,2003,2006-2,2007,2009-1,2004,2006-3,2006-4,2006-5,2005,2008,2010-1}%
, the well-known canonical momenta $p_{\theta}=-i\hbar(\partial_{\theta}%
+\cot\theta/2)$ and $p_{\varphi}=-i\hbar\partial_{\varphi}$ and the momenta
(\ref{deriop1})-(\ref{deriop3}) all seem to be permissible. In fact, neither
of these momenta is all equally physical, nor are they all completely
compatible with Dirac's theory. It is evident that the GM (\ref{GM}) includes
no free parameters so that the momenta (\ref{deriop1})-(\ref{deriop3}) can
never be all true. One purpose of this study is to point out that with the
complete utilization of Dirac's theory, only one case among (\ref{deriop1}%
)-(\ref{deriop3}) survives, which is even fundamental to the orbital angular
momentum therefore the canonical momentum $p_{\varphi}=-i\hbar\partial
_{\varphi}$ turns out to be meaningful, but another canonical momentum
$p_{\theta}=-i\hbar(\partial_{\theta}+\cot\theta/2)$ on the sphere proves to
be problematic in physics. However, as a mathematical symbol without any
physical meaning preassigned,\ the quantity $p_{\theta}$ may still be useful.

This paper is organized as follows. In section II, we present a list of five
different forms of the GM (\ref{deriop1})-(\ref{deriop3}) proposed by
different authors. In section III, we point out that only one form of them is
completely compatible with the framework of Dirac's theory, and thus offers a
proper description of the momentum for the problem under consideration. The
commutator between $p_{\theta}$ and the Hamiltonian $H$ as $[p_{\theta},H]$
breaks Dirac's canonical quantization procedure, meaning that $p_{\theta}$ is
improper. In section IV, it is demonstrated that the GM is fundamental to the
orbital angular momentum, and three components of the GM form the generators
of a dynamic group on the surface. The self-adjointness of the GM is also
discussed in this section. Conclusions and discussions are presented in
section V.

\section{Geometric momenta for a particle on the sphere:\ A review}

For $S^{2}$ parametrized by
\begin{equation}
x=r\sin\theta\cos\varphi,\text{ }y=r\sin\theta\sin\varphi,\text{ }%
z=r\cos\theta, \label{positions}%
\end{equation}
there are five forms of the GM in the current literature, and all can be
written into following form with two real parameters $(\alpha,\beta)$:
\begin{align}
p_{(\alpha,\beta)x}  &  =-i\hbar\frac{\partial}{\partial x}=-\frac{i\hbar}%
{r}(\cos\theta\cos\varphi\frac{\partial}{\partial\theta}-\frac{\sin\varphi
}{\sin\theta}\frac{\partial}{\partial\varphi}-\alpha\sin\theta\cos
\varphi)+\frac{\hbar}{r}\beta\sin\theta\cos\varphi,\label{deriop1}\\
p_{(\alpha,\beta)y}  &  =-i\hbar\frac{\partial}{\partial y}=-\frac{i\hbar}%
{r}(\cos\theta\sin\varphi\frac{\partial}{\partial\theta}+\frac{\cos\varphi
}{\sin\theta}\frac{\partial}{\partial\varphi}-\alpha\sin\theta\sin
\varphi)+\frac{\hbar}{r}\beta\sin\theta\sin\varphi,\label{deriop2}\\
p_{(\alpha,\beta)z}  &  =-i\hbar\frac{\partial}{\partial z}=-\frac{i\hbar}%
{r}(-\sin\theta\frac{\partial}{\partial\theta}-\alpha\cos\theta)+\frac{\hbar
}{r}\beta\cos\theta. \label{deriop3}%
\end{align}
It seems that the corresponding Hamiltonian is then
\begin{equation}
H=\frac{1}{2\mu}(p_{(\alpha,\beta)x}^{2}+p_{(\alpha,\beta)y}^{2}%
+p_{(\alpha,\beta)z}^{2})=\frac{L^{2}}{2\mu r^{2}}+\frac{\hbar^{2}}{2\mu
r^{2}}\gamma, \label{genH}%
\end{equation}
where
\begin{equation}
\gamma=(\alpha-i\beta)(2-\alpha+i\beta). \label{gamma}%
\end{equation}

It is worth stressing that because of the presence of the notorious operator
order problem \cite{kleinert,book5}, there is no simple connection between the
classical Hamiltonian $H=p^{2}/(2\mu)$ and the quantum Hamiltonian
$H=L^{2}/(2\mu r^{2})$. It is therefore understandable that how to identify
the quantum Hamiltonian for particles moving on two-dimensional sphere has
been an intriguing problem. To tackle it, more than four proposals have been
put forward, including two types of group quantization
\cite{higgs,book5,aldaya}, the confining potential procedure \cite{jk,dacosta}%
, and converting the second-class constraints\ into the first class ones or
their equivalent \cite{1998,kleinert,dita,abelian,turkey}. All lead to the
correct one
\begin{equation}
H=\frac{L^{2}}{2\mu r^{2}}. \label{QH}%
\end{equation}
As we will discuss in section III, Dirac's theory itself implies a
self-consistent and insightful way to determine the correct form, but it is
not simply putting parameters $\beta=0$ and $\alpha=0$ or $2$ into
(\ref{genH}) to make it go over to (\ref{QH}). As we see shortly, momenta with
these two sets of parameters are inconsistent with Dirac's theory.

The following is a list of various forms of the GM in the current literature.
Because (\ref{deriop1})-(\ref{deriop3}) give the most general GM of
$p_{(\alpha,\beta)i}$ satisfying the fundamental commutators (\ref{qmc1}
)-(\ref{qmc3}), the differences between them occur in the matter of choice of
the parameters of $(\alpha,\beta)$. The GM was first mentioned in 1968 by
Gyorgyi and Kovesi-Domokos \cite{1968}, who, while investigating the relation
between Schwinger's angular momentum calculus and the Dirac bracket, mentioned
that, on the two-dimensional sphere, the momentum operator $\mathbf{p}$ acting
on the spherical harmonics must take the form $\mathbf{p}=-i\hbar
(\mathbf{\nabla}-\mathbf{n/}r)$ rather than $\mathbf{p}=-i\hbar\mathbf{\nabla
}$ itself. This first appearance of the momentum operator happened to be
complete and it is evidently a special case of the general form (\ref{GM}),
but unfortunately it went unnoticed in all later and relevant studies of the
momentum
\cite{1983,1985,1992,1996,2000,1990,2003,2006-2,2007,2009-1,2004,2006-3,2006-4,2006-5,2005,2008,2010-1}%
.

(i) The first systematic study of the GM was given by Falck and Hirshfeld in
1983, who found $\alpha=$ $2$ and $\beta=0$ \cite{1983}.

(ii) Later in 1985, another form of GM was put forward by Schnitzer who set
$\alpha=$ $0$ or $1$ and $\beta=0$ \cite{1985}.

(iii) In 1992, Ikegami, Nagaoka, Takagi and Tanzawa proposed $\alpha=$ $1$ and
$\beta\neq0$ \cite{1992}.

(iv) In 2000, Hong, Kim and Park suggested that two real parameters ($\alpha$,
$\beta$) should be $\alpha=$ $\pm1$ and $\beta=0$ \cite{2000}.

(v)\ The fifth choice is made by many groups based on different theoretical
grounds and it should be $\alpha=$ $1$ and $\beta=0$
\cite{1968,1990,1996,2003,2006-2,2007,2009-1,2010-1}.

How do we fix the constants $\alpha$ and $\beta$? The primary reason must be a
physical one:\ The two sets of fundamental relation (\ref{qxH}) and
(\ref{pxH}) must be imposed upon the form of GM. As a consequence, we have
only one choice: $\alpha=$ $1$ and $\beta=0$, which coincides with (\ref{GM}),
and $[p_{\theta},H]$ (\ref{pthetaH}) turns out to be inconsistent with the
Dirac's theory. The secondary consideration may come from mathematics:\ None
of these previous approaches (i-v) above have taken the self-adjointness of
the GM into serious consideration, and if considered, only the case with
$\alpha=$ $1$ and $\beta=0$ is satisfactory again. By a self-adjoint operator,
we mean that all its eigenvalues are real and that eigenfunctions
corresponding to distinct eigenvalues are orthogonal. In this sense, the
well-known canonical momentum $p_{\theta}$ is not a self-adjoint operator, as
pointed out many years ago \cite{kato}.

\section{Complete determination of geometric momenta and the Hamiltonian
within Dirac's theory}

For a particle on the sphere with radius $R$ in $R^{3}$, the primary
constraint is $\varphi_{1}$,
\begin{equation}
\varphi_{1}=\vec{x}^{2}-r^{2}\text{ }\approx0; \label{cst1}%
\end{equation}
hereafter the symbol $\approx$ implies a weak equality that is valid on the
constraint surface \cite{note3}. The secondary constraint $\varphi_{2}$ is
given by
\begin{equation}
\varphi_{2}=\vec{x}\cdot\vec{p}\text{ }\approx0. \label{cst2}%
\end{equation}
There are, in fact, third and forth constraint within the consistency from the
calculation of the Poisson bracket $\left\{  \varphi_{j},H_{p}\right\}
_{D}\sim\varphi_{j+1}$ ($j=1,2,3$), where primary Hamiltonian $H_{p}%
=p^{2}/(2\mu)+\lambda(\vec{x}^{2}-r^{2})+\dot{\lambda}p_{\lambda}$, where
$\lambda$ is the Lagrangian multiplier and $p_{\lambda}$ is its conjugate
variable \cite{1983}. The last two constraints $\varphi_{3}={p}^{2}-2\lambda
mr^{2}$\ $\approx0$ and $\varphi_{4}=\dot{\lambda}$\ $\approx0$ are no longer
useful in quantum mechanics \cite{note3}.

In classical mechanics, with the introduction of the Dirac bracket instead of
the Poisson one for the canonical variables $A$ and $B$,
\begin{equation}
\{A,B\}_{D}\equiv\{A,B\}-\{A,\varphi_{\alpha}\}C_{\alpha\beta}^{-1}%
\{\varphi_{\beta},B\}, \label{formu1}%
\end{equation}

\noindent where $\varphi_{\alpha}$ ($\alpha=1,2$) are, respectively, the
primary and second-class constraint (\ref{cst1}) and (\ref{cst2}) and the
matrix elements $C_{\alpha\beta}$ is defined by
\begin{equation}
C_{\alpha\beta}=\{\varphi_{\alpha},\varphi_{\beta}\}, \label{formu2}%
\end{equation}
the primary Hamiltonian $H_{p}$ must then be replaced by the usual
one:\ $H_{p}\rightarrow H,$
\begin{equation}
H=\frac{p^{2}}{2\mu}. \label{CH}%
\end{equation}
The positions ${x}_{{i}}$ and momenta ${p}_{{i}}$ satisfy the following Dirac
brackets,
\begin{equation}
\{{x}_{{i}}{,x}_{j}{\}}_{D}=0,\text{ }\{{x}_{{i}}{,p}_{j}{\}}_{D}=\delta
_{ij}-{\frac{x_{i}x_{j}}{r^{2}}},\text{ }\{{p}_{{i}}{,p}_{j}{\}}_{D}%
=-{\frac{1}{r^{2}}}(x_{i}p_{j}-x_{j}p_{i}), \label{diracb}%
\end{equation}
where other Dirac brackets between ${x}_{{i}}$ and ${p}_{j}$ vanish. The
equation of motion is in general
\begin{equation}
\dot{f}=\{{f},{H\}}_{D}, \label{EOM}%
\end{equation}
from which we have for ${x}_{{i}}$ and $p_{i}$ \cite{note3}
\begin{equation}
\dot{x}=\{{x}_{{i}},{H\}}_{D}=\frac{{p}_{{i}}}{\mu},\text{ }\dot{p}%
_{i}=\{p_{i},{H\}}_{D}=-\frac{x_{i}{p}^{2}}{\mu r^{2}}. \label{cxH}%
\end{equation}

We are ready to construct commutator $[A,B]$ of two variables $A$ and $B$ in
quantum mechanics, which is attainable by direct correspondence of the Dirac
brackets as $[A,B]/(i\hbar)\rightarrow\{A{,B\}}_{D}$. There are evidently two
categories of the fundamental commutator, corresponding to (\ref{diracb})\ and
(\ref{cxH}) respectively. The first category comprises commutator between
operators ${x}_{{i}}$ and ${p}_{{i}}$, given by the quantization of Dirac
brackets (\ref{diracb}):
\begin{align}
\lbrack{x}_{{i}}{,x}_{j}]  &  =0,\label{qmc1}\\
\lbrack{x}_{{i}}{,p}_{j}]  &  =i\hbar(\delta_{ij}-{\frac{x_{i}x_{j}}{r^{2}}%
),}\label{qmc2}\\
\lbrack{p}_{{i}}{,p}_{j}]  &  =-{\frac{i\hbar}{r^{2}}}(x_{i}p_{j}-x_{j}p_{i}).
\label{qmc3}%
\end{align}
There is no operator ordering problem in the right-hand side of
Eq.~(\ref{qmc3}) because the commutator must satisfy the Jacobian identity.
These commutators form a closed algebra, even nonlinear. Given the
two-dimensional spherical surface under consideration, there is a family of
realizations for this algebra, and one can verify that operators
$p_{(\alpha,\beta)i}$ (\ref{deriop1})-(\ref{deriop3}) solve equations
(\ref{qmc1} )-(\ref{qmc3}). The second category of the fundamental commutators
is given by quantization of (\ref{cxH}),
\begin{align}
\lbrack{x}_{{i}}{,H]}  &  =i\hbar\frac{{p}_{{i}}}{\mu},\label{qxH}\\
\lbrack{p}_{{i}}{,H]}  &  =-i\hbar\frac{{x}_{{i}}H+H{x}_{{i}}}{r^{2}}.
\label{pxH}%
\end{align}
Here, in Dirac's quantization from the Dirac bracket $\{p_{i},{H\}}%
_{D}=-2x_{i}{H/}r^{2}$ (\ref{cxH}) to the quantum commutator (\ref{pxH}), we
take the symmetric average of the noncommuting factors $x_{i}$ and ${H}$.
\textit{A complete utilization of Dirac's theory means that operators
}$(x_{{i}},p_{{i}},H)$\textit{\ are necessarily compatible with these two
categories of fundamental commutators (\ref{qmc1})-(\ref{pxH}). }A reasonable
question arises: Can these commutators be sufficient to result in the unique
forms of both the GM and the Hamiltonian? In the general case, the answer is
no because the right-hand side of Eq. (\ref{pxH}) is by no means unique in
quantum mechanics.\ However, the answer to the problem on the two-dimensional
surface is yes. To see this, we start from a quite general form of the
self-adjoint Hamiltonian $H_{q}$ as
\begin{equation}
H_{q}=-\frac{\hbar^{2}}{2\mu}\nabla^{2}+f(\theta,\varphi)\frac{\partial
}{\partial\theta}+g(\theta,\varphi)\frac{\partial}{\partial\varphi}%
+h(\theta,\varphi), \label{hp}%
\end{equation}
where $f(\theta,\varphi),$ $g(\theta,\varphi)$ and $h(\theta,\varphi)$ are
three functions whose explicit forms are to be determined. Substitution of
this $H_{q}$\ into Eq. (\ref{qxH}) results in,
\begin{align}
\lbrack{x,H}_{q}{]}-i\hbar\frac{p_{(\alpha,\beta)x}}{\mu}  &  =(1-\alpha
+i\beta)\frac{\hbar}{\mu}\frac{{x}}{r}+r[-\cos\theta\cos\varphi f(\theta
,\varphi)+\sin\theta\sin\varphi g(\theta,\varphi)],\\
\lbrack y{,{H}_{q}]}-i\hbar\frac{p_{(\alpha,\beta)y}}{\mu}  &  =(1-\alpha
+i\beta)\frac{\hbar}{\mu}\frac{{y}}{r}+r[\cos\theta\sin\varphi f(\theta
,\varphi)+\sin\theta\cos\varphi g(\theta,\varphi)],\\
\lbrack z{,{H}_{q}]}-i\hbar\frac{p_{(\alpha,\beta)z}}{\mu}  &  =(1-\alpha
+i\beta)\frac{\hbar}{\mu}\frac{{z}}{r}+r\sin\theta f(\theta,\varphi).
\end{align}
The terms on the right-hand sides of these equations must be all zero;
otherwise, relation (\ref{qxH}) as $[{x}_{{i}}{,H]}=i\hbar p_{(\alpha,\beta
)i}/\mu$ will be violated. The only solution is simply
\begin{equation}
\alpha=1,\text{ }\beta=0,\text{ }f(\theta,\varphi)=g(\theta,\varphi)=0.
\end{equation}
The only undetermined function $h(\theta,\varphi)$ in ${H}_{q}$ (\ref{hp}) can
be fixed by use of any one of three commutators (\ref{pxH}), and the solution
is
\begin{equation}
h(\theta,\varphi)=0.
\end{equation}
Finally, we not only reproduce the Hamiltonian (\ref{QH}) but also get a
unique choice for the GM that can be simply denoted by ($p_{x}$, $p_{y}$,
$p_{z}$),
\begin{align}
p_{x}  &  =-i\hbar(\cos\theta\cos\varphi\frac{\partial}{\partial\theta}%
-\frac{\sin\varphi}{\sin\theta}\frac{\partial}{\partial\varphi}-\sin\theta
\cos\varphi),\label{hpx}\\
p_{y}  &  =-i\hbar(\cos\theta\sin\varphi\frac{\partial}{\partial\theta}%
+\frac{\cos\varphi}{\sin\theta}\frac{\partial}{\partial\varphi}-\sin\theta
\sin\varphi),\label{hpy}\\
p_{z}  &  =i\hbar(\sin\theta\frac{\partial}{\partial\theta}+\cos\theta).
\label{hpz}%
\end{align}
This offers a proper momentum description for a free particle on a
two-dimensional sphere.

Can a pair of the well-known canonical momenta $(p_{\theta},p_{\varphi})$ be
considered proper? The answer is negative. Application of Dirac's theory to
two angular variables ($\theta,\varphi$) and generalized momenta $(p_{\theta
},p_{\varphi})$ gives rise to nonvanishing relevant Dirac brackets:
\begin{align}
\{\theta,{H\}}_{D}  &  =\frac{p_{\theta}}{\mu r^{2}},\text{ }\{\varphi
,{H\}}_{D}=\frac{p_{\varphi}}{\mu\left(  r\sin\theta\right)  ^{2}}%
,\label{x-H}\\
\{p_{\theta},{H\}}_{D}  &  =\frac{\cot\theta}{\mu\left(  r\sin\theta\right)
^{2}}p_{\varphi}^{2},\text{ \ }\{p_{\varphi},{H\}}_{D}=0. \label{p-H}%
\end{align}
The quantum commutators are simultaneously determined by $\{A{,B\}}_{D}$
$\rightarrow\lbrack A,B]/(i\hbar)$. From two commutators $p_{\theta}=\mu
r^{2}[\theta,{H}]/(i\hbar)$ and $p_{\varphi}=\mu\left(  r\sin\theta\right)
^{2}[\varphi,{H}]/(i\hbar)$, we immediately arrive at the operators
$p_{\theta}=-i\hbar(\partial_{\theta}+\cot\theta/2)$ and $p_{\varphi}%
=-i\hbar\partial_{\varphi}$. Most importantly, we have to know which equations
of motion they obey. The result shows $[p_{\varphi},H]=0$ fulfills all
principles, but another commutator $\left[  p_{\theta},H\right]  $, given by%

\begin{equation}
\left[  p_{\theta},H\right]  =i\hbar\frac{\cot\theta}{\mu\left(  r\sin
\theta\right)  ^{2}}\left(  p_{\varphi}^{2}-\frac{\hbar^{2}}{4}\right)  ,
\label{pthetaH}%
\end{equation}
violates the correspondence: $\left[  p_{\theta},H\right]  (\rightarrow
i\hbar\{p_{\theta},{H\}}_{D})=i\hbar\cot\theta p_{\varphi}^{2}/\mu\left(
r\sin\theta\right)  ^{2}$. So we see that the canonical momentum $p_{\theta}$
breaks Dirac's canonical quantization procedure, and we can safely conclude
that it is not a proper momentum for the free particle on the two-dimensional sphere.

\section{Geometric momenta as the generators of the Lorentz group SO(3,1)}

This section will show that the orbital angular momentum ${L}_{{k}}$
originally defined by $\varepsilon_{ijk}L_{k}\equiv x_{i}p_{j}-x_{j}p_{i}$ is
the consequence of the commutators between momentum components $[{p}_{{i}}%
{,p}_{j}]$, and will give an interpretation of $p^{2}$ in relation
(\ref{genH}) based on the dynamic group. For further demonstration of the fact
that the GM is proper, we will discuss the self-adjointness of the GM
(\ref{hpx})-(\ref{hpz}) at end of the section.

With help of the transformations ${x}_{{i}}\rightarrow{x}_{{i}}/r$ and
$p_{i}\rightarrow{p}_{{i}}/r$, which render the position dimensionless and
cause the momentum to be a dimension of Planck's constant $\hbar$, the
fundamental (nontrivial) commutators\ (\ref{qmc1})-(\ref{qmc3}) become
\begin{align}
\lbrack{x}_{{i}}{,x}_{j}]  &  =0,\label{fc1}\\
\lbrack{x}_{{i}}{,p}_{j}]  &  =i\hbar(\delta_{ij}-x_{i}x_{j}{),}\label{fc2}\\
\lbrack{p}_{{i}}{,p}_{j}]  &  =-i\hbar\varepsilon_{ijk}L_{k}. \label{fc3}%
\end{align}
We can easily verify the following secondary commutation relations:\
\begin{align}
\lbrack{L}_{{i}}{,p}_{j}]  &  =i\hbar\varepsilon_{ijk}{p}_{k},\label{dc1}\\
\lbrack{L}_{{i}}{,x}_{j}]  &  =i\hbar\varepsilon_{ijk}{x}_{k},\label{dc2}\\
\lbrack{L}_{{i}}{,L}_{j}]  &  ={i}\hbar\varepsilon_{ijk}L_{k}. \label{dc3}%
\end{align}
Other commutation relations between ${L}_{{i}}{,x}_{i}$ and ${p}_{i}$ are zero.

Clearly, these commutators (\ref{fc1})-(\ref{dc3}) between ten operators
$\{1,{x}_{{i}},{p}_{{i}},{L}_{{i}}\}$ form a closed algebra. However, it
contains an equation (\ref{fc2}) whose right-hand side is nonlinear in terms
of generator ${x}_{{i}}$. So the algebra is associated with a nonlinear
$SO(3,2)$ group. We leave this nonlinear group for further studies. Moreover,
we can easily identify two familiar subgroups: one is Lorentz group
$SO(3,1)$\ whose six generators are $\{{p}_{{i}},{L}_{{i}}\}$
satisfying\ commutators (\ref{fc3}), (\ref{dc1}), and (\ref{dc3}), and another
is the usual rotational group $SO(3)$ whose three generators are three
Cartesian components of the angular momentum ${L}_{{i}}$, defined by
(\ref{dc3})\ alone. Even the nonlinear $SO(3,2)$ group is a little bit
strange, the Lorentz group $SO(3,1)$ is elementary for theoretical physicists
and the GM ${p}_{{i}}$ acts as the "boost" that a vital part of group
$SO(3,1)$ in our research. Note that the two\ infinitesimal "boosts"
$\delta\psi$ and\ $\delta\phi$ along the $x$ and $y$ directions involve a
rotation $\delta\psi\delta\phi$ about the $z$-axis,
\begin{align}
&  \exp(i\delta\psi{p}_{{x}})\exp(i\delta\phi{p}_{{y}})\exp(-i\delta\psi
{p}_{{x}})\exp(-i\delta\phi{p}_{{y}})\nonumber\\
&  =1-[{p}_{{x}}{,p}_{y}]\delta\psi\delta\phi+...\nonumber\\
&  =1+{i}\hbar{L}_{{z}}\delta\psi\delta\phi+.... \label{boost}%
\end{align}
This implies that three GM operators ${p}_{i}$ are generators of the
infinitesimal parallel transport of a vector on the surface, which will be
studied in detail elsewhere.

Two Casimir operators of Lorentz group $SO(3,1)$ are
\begin{align}
C_{1}  &  \equiv{L}_{{i}}{L}_{{i}}-{p}_{{i}}{p}_{i}=-\hbar^{2}/4,\label{C1}\\
C_{2}  &  \equiv{p}_{i}{L}_{{i}}=0. \label{C2}%
\end{align}
Clearly, quantum Hamiltonian (\ref{QH}) (proportional to ${L}_{{i}}{L}_{{i}}$)
is not the direct quantization of the classical Hamiltonian (\ref{CH})
(proportional to ${p}_{{i}}{p}_{i}$); they differ in a constant that is one of
the Casimir operators of group $SO(3,1)$.

As a consequence of relation (\ref{dc1}), the operators $p_{x}$ and $p_{y}$
can be obtained from $p_{z}$ by rotation of the axis's rotation. Explicitly,
rotation $\pi/2$ around the $y$-axis renders $p_{z}$ to be $p_{x}$, and
$-\pi/2$ around the $x$-axis renders $p_{z}$ to be $p_{y}$,
\begin{equation}
p_{x}=\exp(-i\pi L_{y}/2)p_{z}\exp(i\pi L_{y}/2),\text{ }p_{y}=\exp(i\pi
L_{x}/2)p_{z}\exp(-i\pi L_{x}/2). \label{rotation}%
\end{equation}
Here we follow the convention that a rotation operation affects a physical
system itself \cite{book6}. Hence the eigenvalue problem for operators $p_{x}$
or $p_{y}$ is simultaneously determined once the complete solution to $\hat
{p}_{z}\psi_{p_{z}}(\theta)=p_{z}\psi_{p_{z}}(\theta)$ is known, where over
operator $p_{z}$ on the left-hand side of this equation the carat is used to
distinguish it from the eigenvalue $p_{z}$ on the right-hand side. The
eigenfunctions form a complete set once the eigenvalues $p_{z}$ are real and
continuous,
\begin{equation}
\psi_{p_{z}}(\theta)=\frac{1}{2\pi}\frac{1}{\sin\theta}\tan^{-ip_{z}}\left(
\frac{\theta}{2}\right)  . \label{eigenf}%
\end{equation}
They are $\delta$-function normalized,
\begin{align}
&  \oint{\psi_{{{{{p}^{\prime}}}_{z}}}^{\ast}}\left(  \theta,\phi\right)
{{\psi}_{{{p}_{z}}}}\left(  \theta,\varphi\right)  \sin\theta d\theta
d\varphi\nonumber\\
&  =\frac{1}{2\pi}\int_{0}^{\pi}{\exp\left(  i\left(  {{{{p}^{\prime}}}_{z}%
}-{{p}_{z}}\right)  (\ln\tan\frac{\theta}{2})\right)  }\frac{1}{\sin\theta
}d\theta\nonumber\\
&  =\frac{1}{2\pi}\int_{0}^{\pi}{\exp\left(  i\left(  {{{{p}^{\prime}}}_{z}%
}-{{p}_{z}}\right)  \ln\tan\frac{\theta}{2}\right)  }d\ln\tan\frac{\theta}%
{2}\nonumber\\
&  =\frac{1}{2\pi}\int_{-\infty}^{\infty}{\exp\left(  i\left(  {{{{p}^{\prime
}}}_{z}}-{{p}_{z}}\right)  z\right)  }dz\nonumber\\
&  =\delta\left(  {{{{p}^{\prime}}}_{z}}-{p}_{{z}}\right)  ,
\label{normalization}%
\end{align}
where the variable transformation $\ln\tan\theta/2\rightarrow z$ is used. So,
we see explicitly that the eigenfunctions $\psi_{p_{z}}(\theta)$ form a
complete set. Readers may check that other forms of momentum in (\ref{deriop1}%
)-(\ref{deriop3}) are not the self-adjoint operator.

\section{Conclusions and discussions}

In Dirac's canonical quantization theory on systems with second-class
constraints, the commutators between the {position, momentum} and Hamiltonian
form a set of algebra relations that are fundamental in the construction of
both the quantum momentum and the Hamiltonian. For the particle moving on a
spherical surface or the rotation of a spherical top, it gives rise to
profound consequences. Some of them are in the following. (i) The well-known
canonical momentum operator ${p}_{\theta}$ represents no physical quantities
because it breaks Dirac's canonical quantization procedure, and the proper
momentum is obtainable from an extrinsic description in flat space rather than
an intrinsic one within the surface itself. (ii)\ Three nontrivial commutators
between components of the GM generate three components of the orbital angular
momentum. So another well-known canonical momentum operator ${p}_{\varphi}$ is
meaningful and actually represents the $z$-component of the angular momentum
$L_{z}$. (iii)\ The components of angular momentum and GM are six generators
of the Lorentz $SO(3.1)$ group, where the GM acts as the boost.

The five different forms of the GM in the current literature are revisited,
and only one of them is shown to be self-adjoint and completely compatible
with Dirac's theory. Dirac's canonical quantization theory for the
second-class constraints is more insightful than it used to seem to be, and
one must use it with great care. As far we can see today, among all previous
publications including some by the current authors and possibly this one,
\textquotedblleft Some are correct; some are less correct. Some have original
features; some are less rich in this praiseworthy property\textquotedblright%
\ \cite{book2}. In any case, we hope this contribution to the literature will
be a valuable one.

\begin{acknowledgments}
This work is financially supported by National Natural Science Foundation of
China under Grant No. 11175063, and by the Program for New Century Excellent
Talents in University, Ministry of Education, China. One of the authors (Q. H.
Liu) would like to thank Professor H. J. Zhou for his kind hospitality during
1-21, August 2011, at the Institute of Theoretical Physics (ITP), CAS,
Beijing, where the revision was completed, and for Professors Z. C. Ouyang and
W. M. Zheng at ITP for their kind interest in this work.
\end{acknowledgments}

\_\_\_

{\LARGE About the authors}

\emph{Q. H. Liu}: M., Ph.D. (1996-1999, ITP; supervisiors: Professor H. W.
Peng and Professor Z. C. Ouyang), Professor of Physics (1998-), Director
(2003-) of \textit{School for Theoretical Physics}, Hunan University. Research
interests:\ Quantum mechanics, statistical and condensed matter physics.

\emph{L. H. Tang}:\ M., M.S. candidate (2011, expected; supervisior: Professor
Q. H. Liu).

\emph{D. M. Xun}: F., Ph.D. candidate (2013, expected; supervisior: Professor
Q. H. Liu).


\begin{thebibliography}{99}
\bibitem {note3}As Dirac called, $\varphi_{m}$ $\approx$ $0$ weak equations,
to distinguish them from the usual or strong equations. One can make use of
$\varphi_{m}=0$ only after one has worked out all the Poisson brackets. For
instance, the third constraint $\varphi_{3}={p}^{2}-2\lambda\mu r^{2}$
\ $\approx0$ solves the Lagrangian multiplier $\lambda=$ ${p}^{2}/(2\mu
r^{2})$ which must be used in determining the equation of motion for $\dot
{p}_{i}=[p_{i},H_{p}]=-2\lambda x_{i}=-x_{i}{p}^{2}/(\mu r^{2})$. It is
identical to $\dot{p}_{i}=\left\{  p_{i},H\right\}  _{D}$ where $H=p^{2}%
/(2\mu)$ rather than the primary one $H_{p}=p^{2}/(2\mu)+\lambda(\vec{x}%
^{2}-r^{2})+\dot{\lambda}p_{\lambda}$. In general, with Hamiltonian
$H=p^{2}/(2\mu)$ is used, we must simultaneously use the Dirac bracket in all
cases. In quantum mechanics, there is only $H$ that is free from Lagrangian
multipliers; and the quantum commutators $\left[  A,B\right]  /(i\hbar)$ must
directly correspond to Dirac bracket $\{A,B\}_{D}$.
\end{thebibliography}
\end{document}